\documentclass[prl,twocolumn,showpacs,preprintnumbers,amsmath,amssymb]{revtex4}
\usepackage{graphicx}% Include figure files
\usepackage{dcolumn}% Align table columns on decimal point
\usepackage{bm}% bold math

\def\lsim{\mathrel{\lower2.5pt\vbox{\lineskip=0pt\baselineskip=0pt
\hbox{$<$}\hbox{$\sim$}}}}
\def\gsim{\mathrel{\lower2.5pt\vbox{\lineskip=0pt\baselineskip=0pt
\hbox{$>$}\hbox{$\sim$}}}}

\newcommand{\ima}{{\mbox{Im}\,}}
\newcommand{\rea}{{\mbox{Re}\,}}
\newcommand{\be}{\begin{equation}}
\newcommand{\ee}{\end{equation}}

\newcommand{\NP}[1]{Nucl.\ Phys.\ {#1}}

\newcommand{\PL}[1]{Phys.\ Lett.\ {#1}}

\newcommand{\PR}[1]{Phys.\ Rev.\ {#1}}
\newcommand{\PRL}[1]{Phys.\ Rev.\ Lett.\ {#1}}

\begin{document}

%\preprint{hep ph/0203134}

\title{Nature of the $f_0(600)$ from its
$N_c$ dependence at two loops in unitarized Chiral Perturbation Theory}

\author{J. R. Pel\'aez }
\author{G. R\'{\i}os}

\affiliation{Departamento de F{\'\i}sica Te{\'o}rica II,
  Universidad Complutense de Madrid, 28040   Madrid,\ \ Spain}

%\date{January 2002}% It is always \today, today,

\begin{abstract}
By using unitarized two-loop Chiral Perturbation Theory partial waves
to describe pion-pion scattering
we find that the dominant 
component of the lightest scalar meson does not
follow the  $\bar qq$ dependence on the number of colors
that, in contrast, is obeyed by the lightest vectors. The method 
suggests that a subdominant $\bar{q}q$ component 
of the $f_0(600)$ possibly originates around 1 GeV.
\end{abstract}

\maketitle

%\vspace{-.5cm}
%\section{Introduction}

The lightest scalar mesons 
are a subject of a longstanding controversy that 
is recently receiving relevant contributions
that could help settling the questions
about their existence and nature. 
Experimentally, several analyses \cite{Aitala:2000xu},
find poles for the $f_0(600)$ and $\kappa$,
the lightest scalars with isospin 0
and 1/2, respectively.
The former is of interest for spectroscopy
but also for understanding spontaneous chiral symmetry
breaking, since it has precisely the vacuum
quantum numbers.
On the theoretical side, the QCD
chiral symmetry breaking pattern has been shown to
lead to $f_0(600)$ and $\kappa$ poles in $\pi\pi$ and $\pi K$ scattering
\cite{newsigma,VanBeveren:1986ea,Dobado:1996ps,Oller:1997ti,Oller:1997ng}. Concerning the spectroscopic classification,
the caveat for most hadronic models
is the difficulty to extract the quark
and gluon composition without assumptions hard to justify
within QCD. In contrast, when using fundamental 
degrees of freedom, i.e., in lattice or with QCD inspired potentials,
other complications arise, related to
 chiral symmetry breaking, the use of
actual quark masses or 
the physical pion or kaon masses and decay constants.
All approaches are also complicated by the possible mixing
of different states in the physical one.
Most of these caveats are overcome in a recent approach \cite{Pelaez:2003dy}
(\cite{Pelaez:2004xp} for a review)
based on the pole dependence on the number of colors, $N_c$,
of meson-meson scattering 
within unitarized Chiral Perturbation Theory. 

The relevance of the large $N_c$ expansion \cite{'tHooft:1973jz}
is that it provides an
analytic approximation to QCD in the whole
energy region and a
clear identification of
$\bar qq$ states, that become
bound states as $N_c\rightarrow\infty$,
and whose masses scale as $O(1)$ and 
their widths as $O(1/N_c)$. Other kind of hadronic 
states may show different behaviors \cite{Jaffe}.

In order to avoid any spurious $N_c$ dependence in the
hadronic description, we use Chiral Perturbation Theory (ChPT),
which is the QCD low energy Effective Theory, and where
the large $N_c$ behavior is implemented systematically.
It is built as 
the most general derivative expansion of a Lagrangian \cite{chpt1},
in terms of $\pi, K$ and $\eta$ mesons compatible with the QCD symmetries,
These particles 
are the Goldstone bosons associated to the spontaneous
chiral symmetry breaking of massless QCD and 
are therefore the lightest degrees of freedom.
Actually, the $u$, $d$ and $s$ quark
masses are non-vanishing but small enough to be treated as perturbations
that give rise to $\pi, K$ and $\eta$ masses.
Thus, ChPT is an expansion in powers
of momenta and masses and,
generically, its applicability
is limited to a few hundred MeV above threshold.
Each order is made of all possible
terms multiplied by a ``chiral'' parameter.
These Low Energy Constants (LECS) are renormalized
to absorb loop divergences order by order, and once 
determined from experiment they can be used
in any other pseudo-Goldstone boson amplitude.

For our purposes, we are interested in 
meson-meson scattering amplitudes, since 
by unitarization they generate dynamically resonances 
not initially present in ChPT 
\cite{Truong:1988zp,Dobado:1996ps,Oller:1997ng,Guerrero:1998ei,GomezNicola:2001as,Oller:1997ti}.
Indeed \cite{GomezNicola:2001as,Pelaez:2004xp}, 
with the coupled channel Inverse Amplitude Method (IAM),
the one loop ChPT meson-meson amplitudes describe data up to  
roughly $\sqrt{s}\simeq1.2\,$GeV and generate the $\rho$ and $K^*$ vectors, 
as well as the $f_0(600)$, $\kappa$, $a_0(980)$ and $f_0(980)$ 
scalars, and, most importantly, using LECS
compatible with standard ChPT and therefore
{\it without any further assumption or source of spurious $N_c$ dependence}. 

By scaling the one-loop ChPT parameters with their $N_c$ behavior,  
it was recently shown that the generated $\rho$ and $K^*$ show the
typical $N_c$ behavior of $\bar{q}q$ states, whereas the scalars  are at odds 
with a dominant $\bar{q}q$ component. 
These results, confirmed 
by other methods \cite{Uehara:2003ax}, implied some cancellation
between tree level diagrams proportional to LECS, and that 
loop diagrams with two intermediate mesons
are very relevant in the generation of light scalars.
But such loop diagrams are subdominant in the large $N_c$ counting
and one could wonder about  
the stability under small changes in the LECS and
about higher order ChPT corrections that could become larger than
loop terms at sufficiently large $N_c$, and reveal
the existence of subdominant $\bar{q}q$ components.

Here we present a method to quantify the
above statements,
and generalize the approach of \cite{Pelaez:2003dy}
to two-loop ChPT and in particular to $\pi\pi$
scattering \cite{Bijnens:1997vq}. 
Despite the many second order parameters and 
their large uncertainties, the data can be well described
and we find once more that the $\rho(770)$
behaves as $\bar{q}q$ with $N_c$, whereas
the $f_0(600)$ main component 
does not behave as such. 
%Even more, it is not possible to impose
%a dominant $\bar{q}q$ behavior on the $f_0(600)$ and the $\rho$ simultaneously, 
%but the second order result suggests a subdominant $\bar{q}q$
%$f_0(600)$ component around $1$ GeV arising at larger $N_c$.
Furthermore, with the second order calculation
a dominant $\bar{q}q$ behavior cannot
be imposed on the $f_0(600)$ and the $\rho$ simultaneously, 
but a subdominant $\bar{q}q$
$f_0(600)$ component  seems to arise at larger $N_c$ around $1$ GeV.

Thus, at leading order, the only parameter is
the pion decay constant in the chiral limit, $f_0=O(\sqrt{N_c})$, fixed by the 
spontaneous symmetry breaking scale $4\pi f_0\simeq1\,$GeV. 
Indeed, ChPT $\pi\pi$ scattering amplitudes are expanded as
$t\simeq t_2+t_4+t_6+\,\ldots$ with $t_k=O((p/4\pi f_0)^k)$
and are, generically, $O(1/N_c)$.
In particular, the LECS appearing in $\pi\pi$ scattering at $O(p^4)$ \cite{chpt1},
all scale as
$O(N_c)$. For simplicity we use
the $SU(2)$ notation, $l_i^r$, since we are 
only dealing with $\pi\pi$ scattering 
(see the last reference in \cite{chpt1} for a translation to $SU(3)$).
In Table I we give a sample of 
$l_i^r$ sets given in the literature, whose differences
we take as systematic uncertainties for the set we use in our fits below.
In Table II we also list the six 
$O(p^6)$ constants that appear in 
$\pi\pi$ scattering, denoted $r_i$. They all count as $O(N_c^ 2)$.
Those values are just {\it estimates}
assuming they are saturated by the multiplets of 
the lightest (predominantly vector) resonances. 
This hypothesis works well at $O(p^ 4)$
\cite{Ecker:1988te},
but for $O(p^ 6)$ is probably just correct within an order of magnitude
\cite{Bijnens:1997vq} and we conservatively assign a 100\% uncertainty.

\begin{table}
\begin{tabular}{|c|c|c|c|c|c|c|c|c|}
\hline
& \multicolumn{4}{c|}{$O(p^4)$ LECS} & \multicolumn{4}{c|}{$O(p^6)$ LECS}\\ 
\cline{2-9}
%& Ref & Ref & Ref & At $O(p^4)$ %Ref 
%& Ref & Ref &&At $O(p^6) $ \\
Refs. &\cite{chpt1}&\cite{BijnensGasser}&\cite{Pelaez:2004xp}&
we use%$\chi^2_{LECS}$
&\cite{Amoros:1999qq}&\cite{Bijnens:1997vq}&\cite{Girlanda:1997ed}& 
we use%$\chi^2_{LECS}$ 
\\
\hline
$10^3 l_1^r$& -6.0 & -5.4 & -3.5 & 3.5$\pm$2.2  &-3.3  & -5.2  & -4.6  &-3.3$\pm$2.2 \\
$10^3 l_2^r$& 5.5  & 5.7  &  4.7 & 4.7$\pm$1.0  & 2.9  &  2.3  &  2.0  &2.9$\pm$1.0 \\
$10^3 l_3^r$& 0.82 & 0.82 & -2.6 & 0.82$\pm$3.8 & 1.2  &  0.82 &  0.82 &0.82$\pm$3.8 \\
$10^3 l_4^r$& 5.6  & 5.6  &  8.6  & 6.2$\pm$2.0 & 2.4  &  5.6  &  6.2  &6.2$\pm$2.0 \\
\hline
\end{tabular}
\caption{Sample of LECS central values. The fifth and ninth
columns, whose uncertainties roughly represent
the previous sample, are used in our fits in the text 
to calculate $\chi^2_{LECS}$.}
\end{table}

The large $N_c$ counting does not specify at what renormalization scale
$\mu$ it applies, thus becoming an
uncertainty studied in \cite{Pelaez:2003dy} for the one-loop LECS.
For the $r_i$, the scale dependence is much more cumbersome 
and has not been written explicitly. Nevertheless, in both cases
it is subleading in $1/N_c$, and given the fact that we have a 100\% 
error on the $r_i$ should
be well within errors for our fits. Hence,
we do not perform such analysis here, simply setting $\mu=770\,$MeV,
as usual. 

Next, resonances can be found as poles in partial wave amplitudes
$t_{IJ}$ of isospin $I$ and angular momentum $J$ that,
in the elastic regime satisfy the unitarity condition:
\begin{equation}
  \ima t =\sigma \vert t\vert^2 \Rightarrow 
\ima \frac{1}{t}=-\sigma \Rightarrow
t=\frac{1}{\rea t^{-1} - i \sigma},
\label{unit}
\end{equation}
where $\sigma$ is the known two-meson phase space
and we have omitted the $IJ$ indices for brevity.
Note that ChPT expansions violate {\it exact} unitarity,
since in the first Eq.(\ref{unit}),
 the highest power of momenta on the right hand 
is twice that on the left. Unitarity is only satisfied
{\it perturbatively}
\begin{equation}
\ima t_2 = 0,\quad \ima t_4 = \sigma t_2^2,\quad
\ima t_6= \sigma 2 t_2 \rea t_4, \;\ldots
\label{pertunit}
\end{equation}
If we replace in Eq.(\ref{unit}) $\rea t^{-1}$ by its
ChPT approximation we get the Inverse Amplitude Method (IAM),
that satisfies elastic unitarity exactly.
At $O(p^4)$ it reads, 
\begin{equation}
  \label{eq:IAM4}
%  t\simeq\frac{t_2^2}{t_2-t_4},
  t\simeq t_2^2/(t_2-t_4),
\end{equation}
and its fit to ``data only'' is 
listed in  Table II, in the SU(2) notation.
The fit quality is remarkable, given
the huge systematic uncertainties, (conservatively $\pm5^0$ and 5\% error
for the $f_0(600)$ channel) and we refer to
\cite{GomezNicola:2001as,Pelaez:2004xp}
for details and figures with a comparison with data.
Using Eqs.(\ref{unit}) and (\ref{pertunit}),
the $O(p^6)$ IAM \cite{Dobado:1996ps,Nieves:2001de} reads
\begin{equation}
  \label{eq:IAM6}
%  t\simeq\frac{t_2^2}{t_2-t_4+t_4^2/t_2-t_6},
t\simeq t_2^2/(t_2-t_4+t_4^2/t_2-t_6),
\end{equation}
that recovers the $O(p^6)$ ChPT expansion 
at low energies and describes well elastic $\pi\pi$ scattering 
data \cite{Nieves:2001de}. 
In addition, the IAM has a right cut that defines two Riemann sheets. In the second
sheet we find poles associated to resonances; in particular, 
for the $\rho(770)$ in the $(I,J)=(1,1)$ channel and 
for the $f_0(600)$
in the $(0,0)$ one. For narrow resonances, $\Gamma<<M$,
the pole position is related
to its mass and width as $\sqrt{s_{\rm pole}}\simeq M- i \Gamma/2$,
and we keep this as a {\it definition}
for the wide $f_0(600)$, whose $M\sim400-500\,$MeV
and $\Gamma\sim400-600\,$MeV.

By scaling the previous
parameters with their dominant $N_c$ behavior, namely,
$f_{N_c}\rightarrow f \sqrt{N_c/3}$, 
$l^r_{i,\,N_c}\rightarrow l_i^r N_c/3$ and $r_{i,\,N_c}\rightarrow r_i (N_c/3)^2$,
we obtain the large $N_c$ dependence of
$M_{N_c}$ and $\Gamma_{N_c}$ of the $\rho$ and $f_0(600)$ poles 
generated by the IAM.
If a resonance is predominantly a $\bar{q}q$, 
$M_N\sim O(1)$ and $\Gamma_N\sim O(1/N_c)$, and 
so  it was shown \cite{Pelaez:2003dy} that the $O(p^4)$ IAM
reproduced remarkably well that behavior for the 
$\rho(770)$ and $K^*(892)$, two well established $\bar{q}q$ mesons.
This is the expected behavior if in Eq.(\ref{eq:IAM4})
one neglects the two-meson loop terms, 
which are subleading at large $N_c$ with respect to $O(p^4)$ LECS contributions.

In contrast, the lightest scalars follow
a qualitatively different behavior. Loop diagrams, 
instead of the $O(p^4)$ LECS terms, play a 
relevant role in determining the scalar pole position.
This is nothing but the well 
known fact that light scalars are dynamically generated
by the resummation in Eq.(\ref{eq:IAM4})
of two-meson loop diagrams \cite{Dobado:1996ps,Guerrero:1998ei,GomezNicola:2001as,Oller:1997ti,Oller:1997ng}. 
However, although relevant at $N_c=3$, loop diagrams are suppressed
by $1/N_c$ compared to tree level
terms with LECS, and the $O(p^ 6)$ terms could become bigger
at some larger $N_c$, where the $O(p^ 4)$ $N_c$ results should no
longer be trusted. 
For that reason it is important to
check the $O(p^ 6)$ IAM: it should give small
corrections to the $O(p^ 4)$ close to $N_c=3$, but it may deviate
at larger $N_c$ and even unveil some subdominant $\bar qq$ component.

Still, before scaling the $O(p^6)$ IAM,
let us first note that $M_{N_c}=O(1)$ and $\Gamma_{N_c}=O(1/N_c)$
is only the  {\it leading} $\bar{q}q$ scaling. Taking into account
subleading uncertainties, to consider a resonance 
a $\bar{q}q$ state, it is enough that 
\begin{equation}
M_{N_c}^{\bar{q}q}=\widetilde{M} \left(1+\frac{\epsilon_M}{N_c}\right),\;
\Gamma_{N_c}^{\bar{q}q}=\frac{\widetilde{\Gamma}}{N_c}
\left(1+\frac{\epsilon_\Gamma}{N_c}\right),
\label{ncqqpoles}
\end{equation}
were $\widetilde{M}$ and $\widetilde{\Gamma}$ are unknown but $N_c$ independent
and the subleading terms have been gathered 
in $\epsilon_M, \epsilon_\Gamma$, which are $O(1)$.
Thus, for a $\bar{q}q$ state, 
the {\it expected} $M_{N_c}$ and $\Gamma_{N_c}$ can be obtained
from those at $N_c-1$ generated by the IAM,
\begin{eqnarray}
M_{N_c}^{\bar{q}q}&\simeq&M_{N_c-1}\left[1+\epsilon_M\left(\frac1{N_c}-\frac1{N_c-1}\right)\right]\\\nonumber
&\equiv&M_{N_c-1}+\Delta M^{\bar{q}q}_{N_c}, \\
\Gamma_{N_c}^{\bar{q}q}&\simeq&\frac{\Gamma_{N_c-1}\,(N_c-1)}{N_c}\left[ 1+
\epsilon_\Gamma \left(\frac1{N_c}-\frac1{N_c-1}\right)\right]\\\nonumber
&\equiv& \frac{\Gamma_{N_c-1}\,(N_c-1)}{N_c}+\Delta\Gamma_{N_c}^{\bar{q}q}.
\end{eqnarray}
Note the $\bar{q}q$ index for all quantities
obtained {\it assuming} a $\bar{q}q$ behavior.
We refer the values at $N_c$ to those at $N_c-1$ because then
we will be able to  calculate from what $N_c$ value a resonance 
starts behaving as a $\bar{q}q$, which is of interest
%if the physical state is a mixture of different components at $N_c=3$.
in order to look for subdominant $\bar qq$ components.
Thus, we can now define an {\it averaged} 
$\bar\chi_{\bar{q}q}^2$%=\chi^{2}_M+\chi^{2}_\Gamma$ 
to measure how close a resonance is to 
a $\bar{q}q$ behavior,
 using as uncertainty the  $\Delta M^{\bar{q}q}_{N_c}$ and
$\Delta\Gamma_{N_c}^{\bar{q}q}$. 
\begin{equation}
  \label{eq:avchi}
  \bar\chi_{\bar{q}q}^2\!=\!\frac1{2n}\sum_{N_c=4}^{n}\!\left[
\left(\frac{M_{N_c}^{\bar{q}q}-M_{N_c}}
{\Delta M_{N_c}^{\bar{q}q}}
\right)^{\!\!2}
\!\!+
\left(\frac{\Gamma_{N_c}^{\bar{q}q}-\Gamma_{N_c}}
{\Delta\Gamma_{N_c}^{\bar{q}q}}\right)^{\!\!2}
\right]
\end{equation}

Since at $N_c=3$ 
we expect {\it generically} 30\% uncertainties
we take $\epsilon_M=\epsilon_\Gamma=1$.
Let us note that $\Delta M$, and even faster $\Delta\Gamma$, tend to zero 
for large $N_c$ and eventually become smaller than our precision 
determining the pole position, 1 MeV, which we add as a systematic error. 
When a state is predominantly $\bar{q}q$, it should follow
Eq.(\ref{ncqqpoles}) and $\bar\chi_{\bar{q}q}^2\lsim1$. Otherwise $\bar\chi_{\bar{q}q}^2\gg1$.
Note that $n$ should not be too far from 3, since we are looking
for the $N_c$ behavior of the physical state. If we took $n$ too large,
we could be changing radically the original mixture of the observed state
and for sufficiently large $N_c$ even the tiniest $\bar{q}q$ component
could become dominant over the rest \cite{Sun:2005uk}. Therefore, our method
first determines the behavior of the resonance {\it dominant component},
but also, when $\bar{q}q$ is not dominant, the $N_c$ at which it becomes so.

 Furthermore,
by minimizing its $\bar\chi_{\bar{q}q}^2$ we can constrain 
a state to follow the $\bar{q}q$ behavior. 
Thus, we will minimize $\chi^2_{data}+\chi^2_{LECS}$
plus the $\bar\chi_{\bar{q}q}^2$ of either of the 
$\rho$, or the $f_0(600)$, or both.
The averaged $\chi^2_{LECS}$ measures how far the fitted
LECS are from their typical values in the tables
and stabilizes them. Note that
$\pi\pi$ scattering data are poor, with large 
systematic uncertainties and 
not very sensitive to some of the individual 
parameters 
but to their combinations, thus producing large correlations,
driving the LECS away from the typical values
for tiny improvements in the $\chi^2$,
particularly at $O(p^6)$.
We will provide the $\chi^2_{data}$, $\chi^2_{LECS}$, and
$\bar\chi_{\bar{q}q}^2$, divided by the number of data points, 
the number of LECS and $2n$, respectively.

Thus, in Table II we show
three $O(p^4)$ fits: 
to data only, constrained to a $\rho$ $\bar{q}q$ hypothesis,
or constraining the $f_0(600)$ to be a $\bar{q}q$.
We list for each fit
the different $\chi^2$ described above and
we see that our approach clearly identifies
the $\rho$ as a $\bar{q}q$, since 
$\bar \chi_{\rho, \bar{q}q}^2\simeq0.25$.
In contrast, $\bar \chi_{\bar{q}q}^2\geq125$ for the $f_0(600)$, 
even if we constrain the fit to minimize also $\bar \chi_{\bar q q}^2$
for the $f_0(600)$ (at the price of a higher $\chi^2_{LECS}=5.6$).
This is the quantitative statement of the  $O(p^4)$ results in \cite{Pelaez:2003dy,Pelaez:2004xp} where it was concluded that 
the main component of the $f_0(600)$
was not $\bar{q}q$.

Unfortunately, the  $O(p^6)$ analysis has a
large freedom and thus $\chi^2_{LECS}$
plays a relevant role to stabilize the fit, 
but keeping in mind that the $r_i$
uncertainties were arbitrarily chosen to be 100\%.
In Table II and Fig.~1
we show three $O(p^6)$ fits: constraining the $\rho$
as a $\bar qq$ (Fig.~1. Top), or the $f_0(600)$ (Fig.~1. Center)
or both (Fig.~1, Bottom).
As expected, the $O(p^6)$ results are consistent with 
those at $O(p^4)$ not far from $N_c=3$ \cite{Pelaez:2003dy,Pelaez:2004xp} but
for the scalar channel they deviate around $N_c\sim8$.

In particular, in the ``$\rho$ as $\bar{q}q$'' fit a $\bar{q}q$ dominant nature
comes out neatly for the $\rho$, whose $\bar \chi_{\bar{q}q}^2\sim0.9$,
but is discarded
for the $f_0(600)$, since its $\bar \chi_{\bar{q}q}^2\simeq15$
and Fig.1 shows that its mass and width
{\it both rise} when $N_c$ increases {\it not too far from real life, $N_c=3$}.
However, for $N_c>8$ 
\emph{the mass tends to a constant around 1 GeV and
the width decreases}, but not
with a $1/N_c$ scaling. \emph{This suggests
a mixing with a $\bar qq$ subdominant component},
arising as loop-diagrams become
more suppressed at large $N_c$.

\begin{table}
\begin{tabular}{|c|c|c|c|c|c|c|}
\hline
& \multicolumn{3}{c|}{fits $O(p^4)$} & \multicolumn{3}{c|}{fits $O(p^6)$} \\ \cline{2-7}
Fit & Only & $\rho$ as & $f_0(600)$ & $\rho$ as &$f_0(600)$ & $\rho,f_0(600)$ \\
& data &$\bar{q}q$ & as $\bar{q}q$ & $\bar{q}q$ & as $\bar{q}q$ &as $\bar{q}q$ \\ \hline
$10^3l_1$&-3.8&-3.8&-3.9 &-5.4 &-5.7 &-5.7  \\
$10^3l_2$&4.9&5.0&4.6    & 1.8 & 2.6 & 2.5  \\
$10^3l_3$&0.43&0.42&2.6  & 1.5 &-1.7 & 0.39 \\
$10^3l_4$&7.2&6.4&15     & 9.0 & 1.7 & 3.5  \\
\hline
$\chi^2_{data}$ &\bf 1.1 &\bf 1.2 &\bf 1.4 &\bf 1.1 &\bf 1.4 &\bf 1.5 \\
$\chi^2_{LECS}$ &\bf 0.08 &\bf 0.03 &\bf 5.6 &\bf 1.9 &\bf 2.1 &\bf 1.4  \\
$\chi^2_{\rho,\bar{q}q}$ &0.26 &\bf 0.22 &0.32 &\bf 0.93 &2.0 &\bf 1.3 \\
$\chi^2_{f_0(600),\bar{q}q}$ &140 & 143 &\bf 125 &15 &\bf 3.5 &\bf 4.0 \\ \hline
$10^4r_1$ &                      &-0.6 &         &-0.60 &-0.60 &-0.58 \\
$10^4r_2$ &                     &1.3  &Our      & 1.5  & 1.3  & 1.5  \\
$10^4r_3$ & Ref.                 &-1.7 &$O(p^6)$ &-1.4  &-4.4  &-3.2  \\
$10^4r_4$ &\cite{Bijnens:1997vq} &-1.0 &fits     & 1.4  &-0.03  &-0.49 \\
$10^4r_5$ &                      &1.1  &         & 2.4  & 2.7  & 2.7  \\
$10^4r_6$ &                      &0.3  &         &-0.60 &-0.70 &-0.62 \\ \hline
\end{tabular}
\caption{ IAM fits to data or constrained to a $\bar qq$ $N_c$ behavior
for the $\rho$ and $f_0(600)$. In boldface the $\chi^2$ minimized on each fit.
For the $O(p^6)$ fits we also provide
the $r_i$ estimates \cite{Bijnens:1997vq}
}
\end{table}

One might wonder if the $f_0(600)$ could also be forced to behave 
predominantly as a $\bar{q}q$.
Thus we made a  ``$f_0(600)$ as a $\bar qq$'' constrained
fit 
(Fig.~1, Center). 
The price to pay is a deterioration of the $\chi^2_{data}$ and
an unacceptable 
$\rho$ $\bar qq$ behavior, since  its $\bar \chi_{\bar{q}q}^2\sim 2$.
Still, the $f_0(600)$  
$\bar \chi_{\bar{q}q}^2$ decreases only to $3.5$
(for 34 $N_c$ points). This extreme case allows us to conclude 
that \emph{the $O(p^6)$ calculation
cannot accommodate a $\bar{q}q$ dominant component for the $f_0(600)$. }

Finally, we have studied how much of a \emph{subdominant} $\bar qq$
behavior the $f_0(600)$ can accommodate without spoiling 
that of the $\rho$.
Hence,
we have also minimized in the fit
 the  $\bar \chi_{\bar{q}q}^2$ both for the $\rho$ and $f_0(600)$
(Fig.~1, Bottom).
The $f_0(600)$ still does not behave predominantly as a $\bar q q$,
since its $\bar \chi_{\bar{q}q}^2\simeq4$.
However, it starts behaving as 
a $\bar q q$, i.e., $\bar \chi_{\bar{q}q}^2\simeq1$,
for $N_c\geq6$. The $\bar qq$ behavior of the $\rho$
only deteriorates a little, $\bar \chi_{\bar{q}q}^2\simeq1.3$, and should
not be pushed much further.
This result suggests that the subdominant mixing with a $\bar qq$ state
around 1 GeV seen in the first fit,
would become dominant around $N_c>6$, {\it at best} .

\begin{figure}%[h]

\hbox{
\includegraphics[scale=.37,angle=-90]{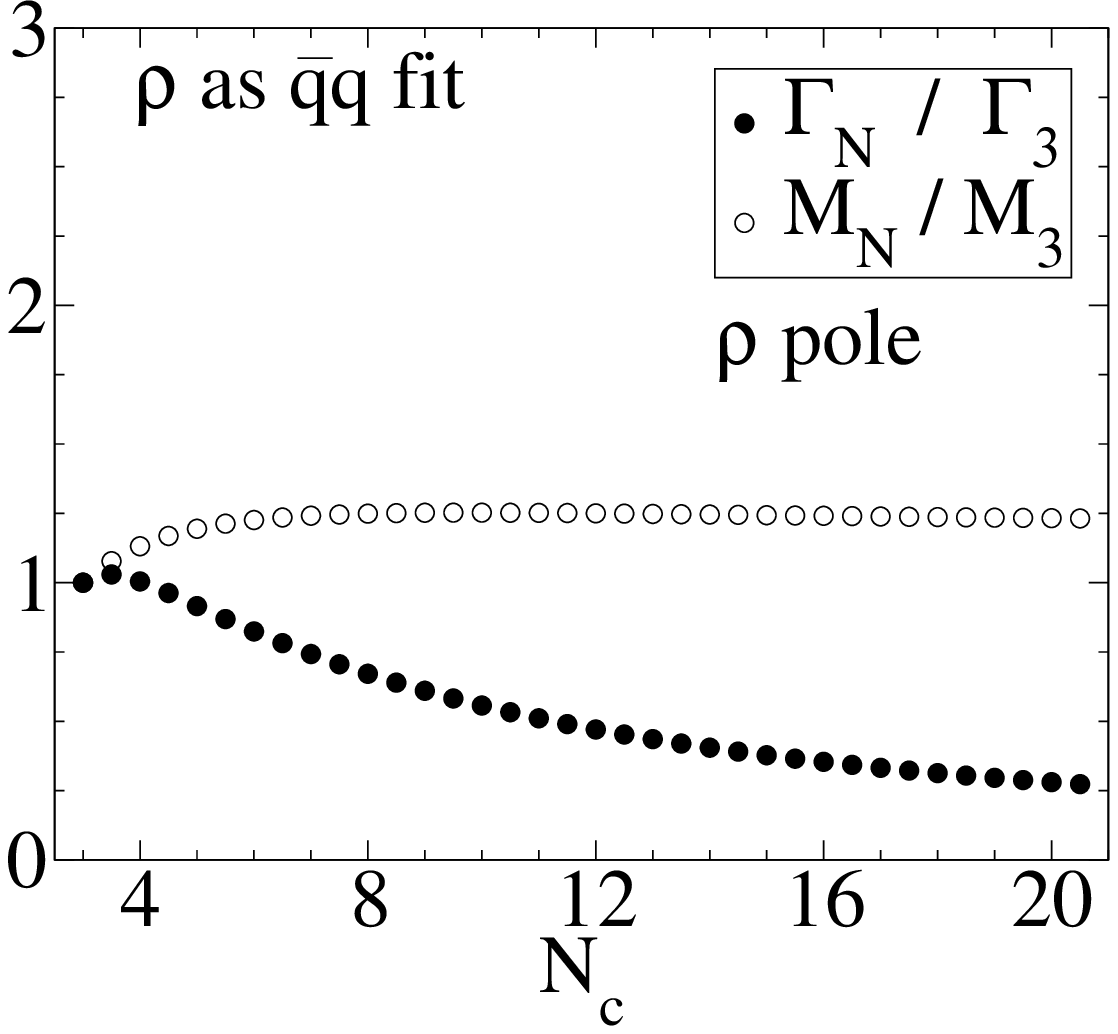}% Here is how to import EPS art
%\caption{\rm Ejemplo figura}
%\end{figure}
%\begin{figure}[h]
\includegraphics[scale=.37,angle=-90]{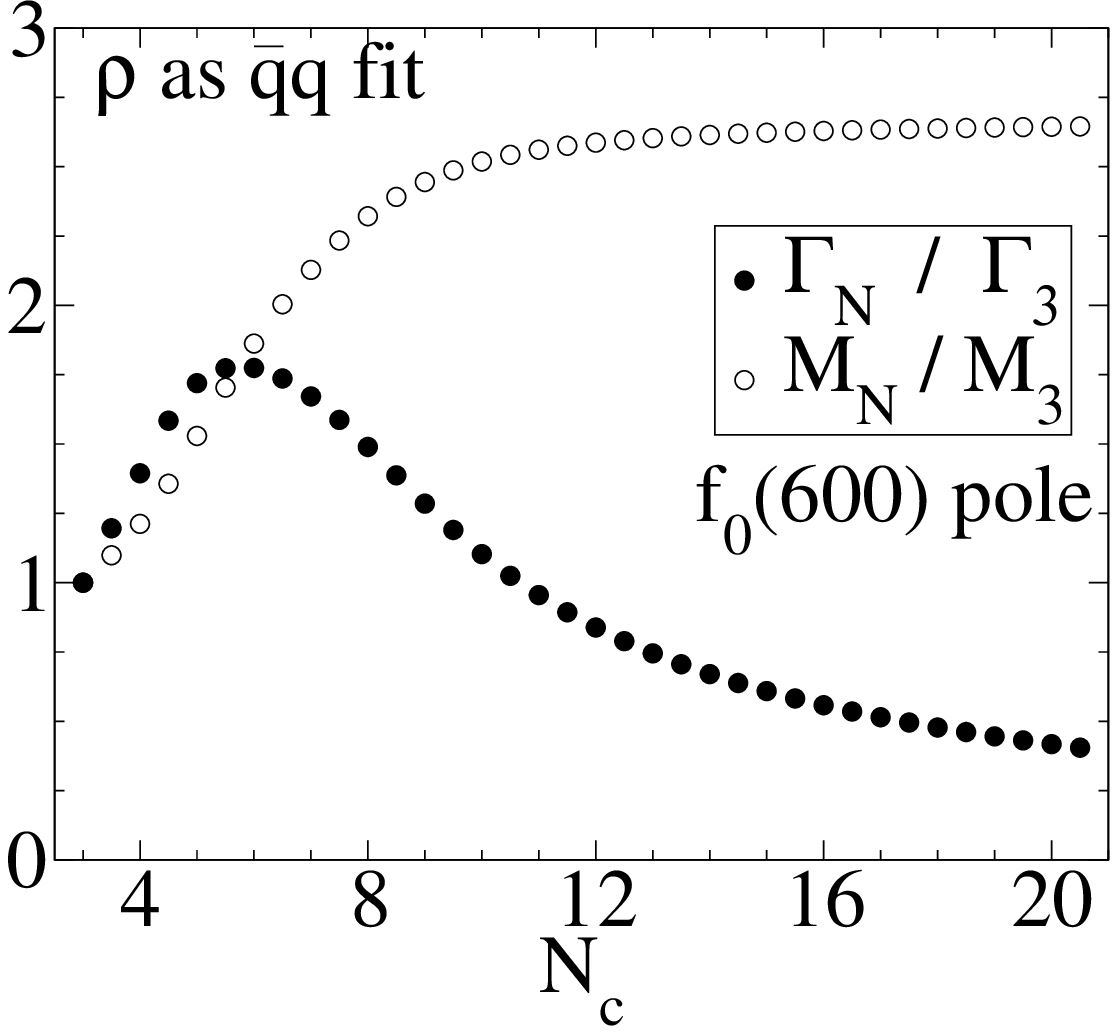}% Here is how to import EPS art
}

\hbox{
\includegraphics[scale=.37,angle=-90]{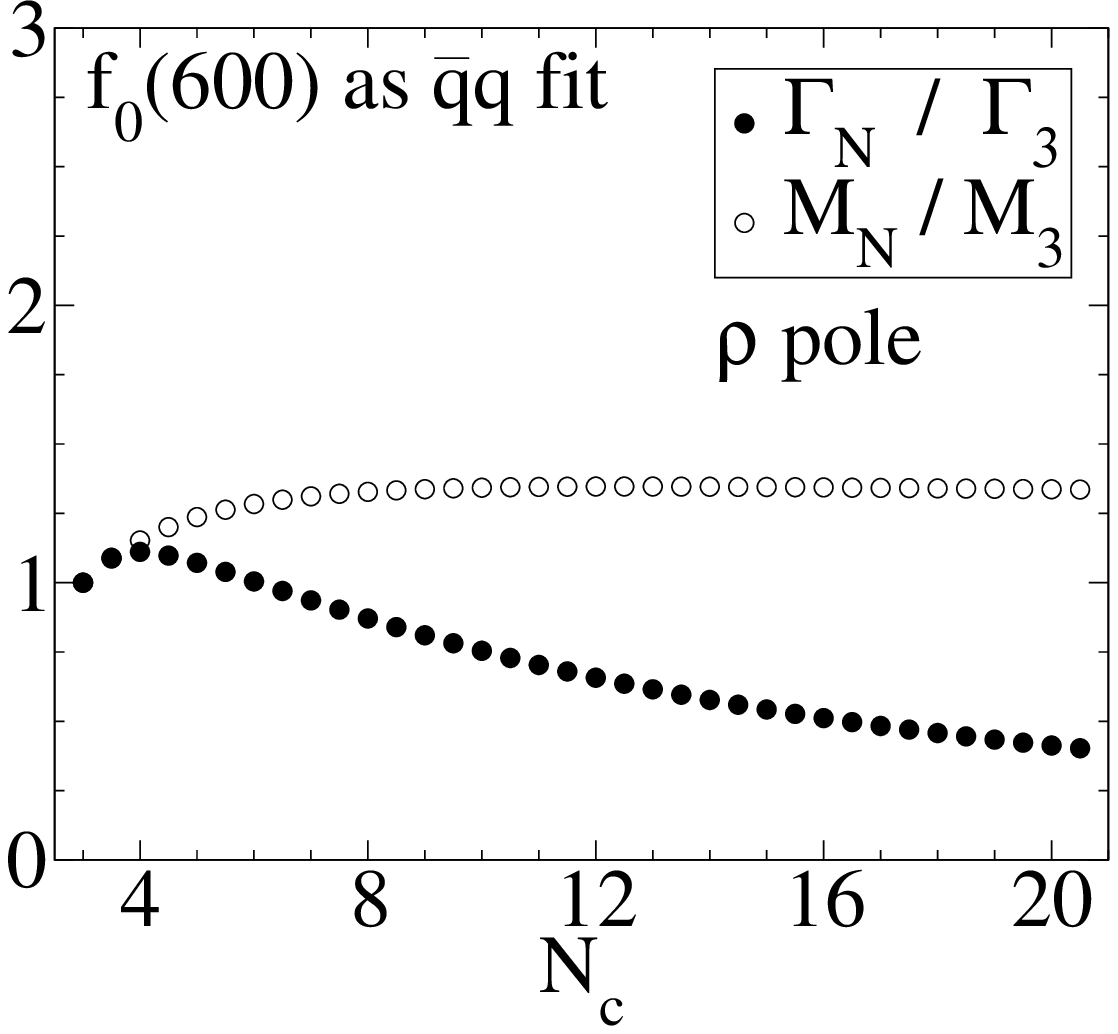}% Here is how to import EPS art
%\caption{\rm Ejemplo figura}
%\end{figure}
%\begin{figure}[h]
\includegraphics[scale=.37,angle=-90]{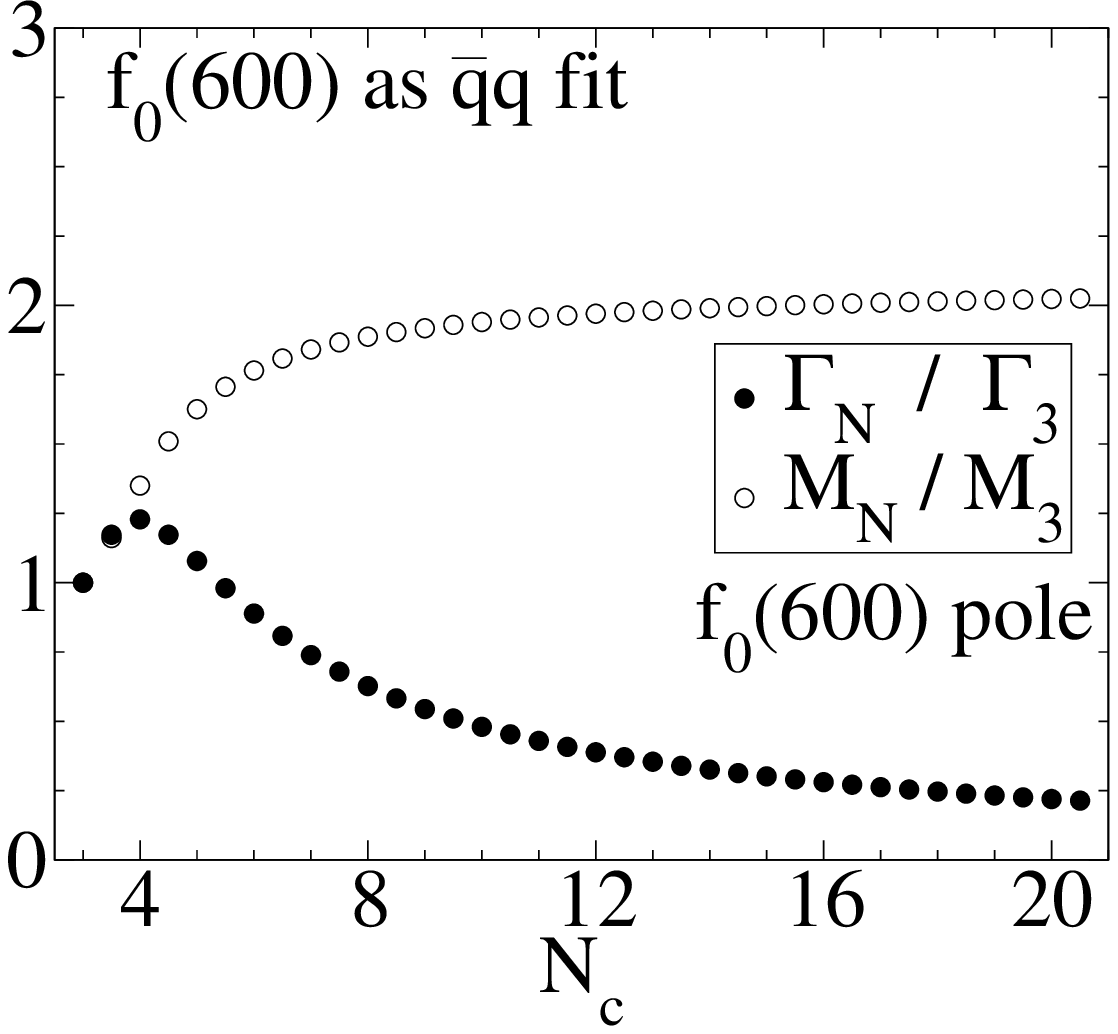}% Here is how to import EPS art
}

\hbox{\includegraphics[scale=.37,angle=-90]{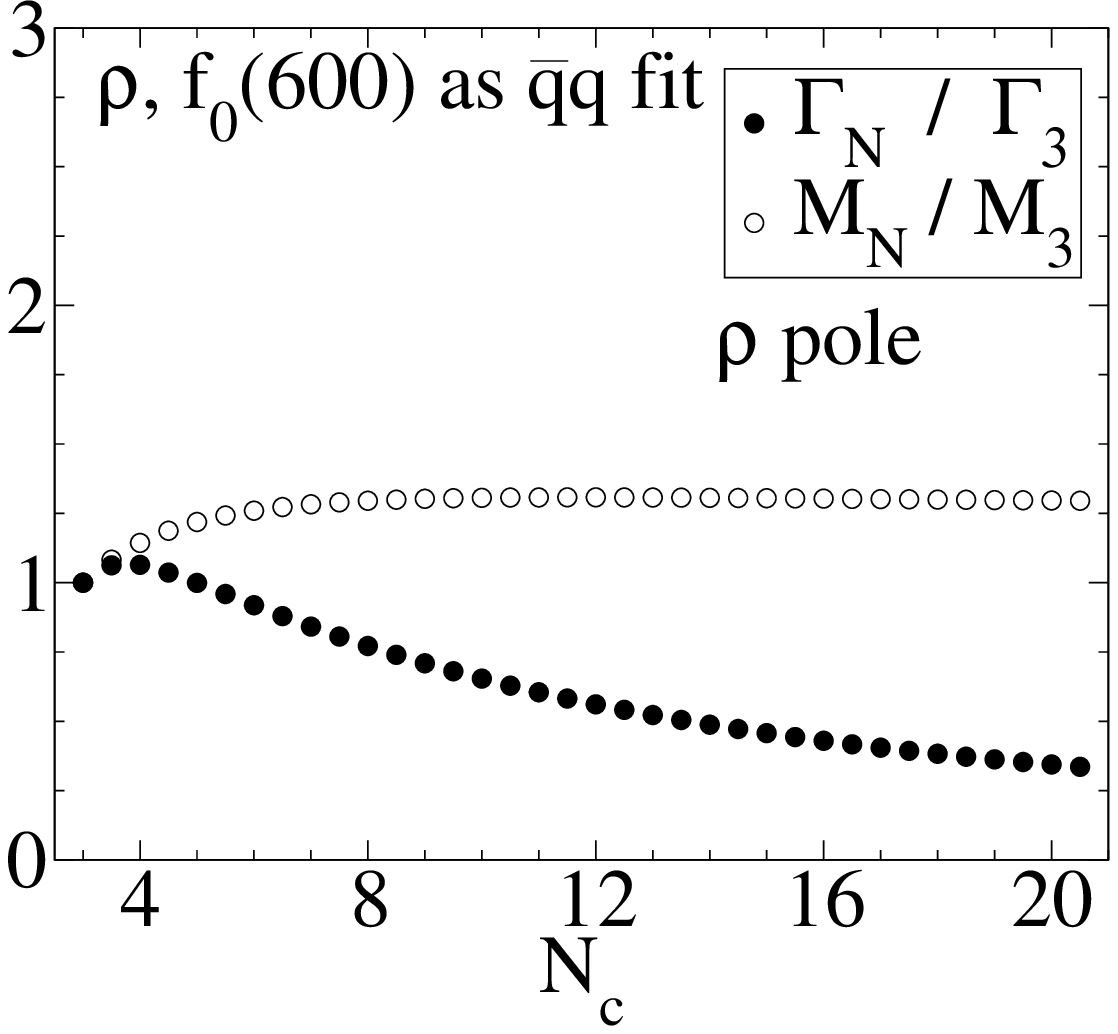}% Here is how to import EPS art
%\caption{\rm Ejemplo figura}
%\end{figure}
%\begin{figure}[h]
\includegraphics[scale=.37,angle=-90]{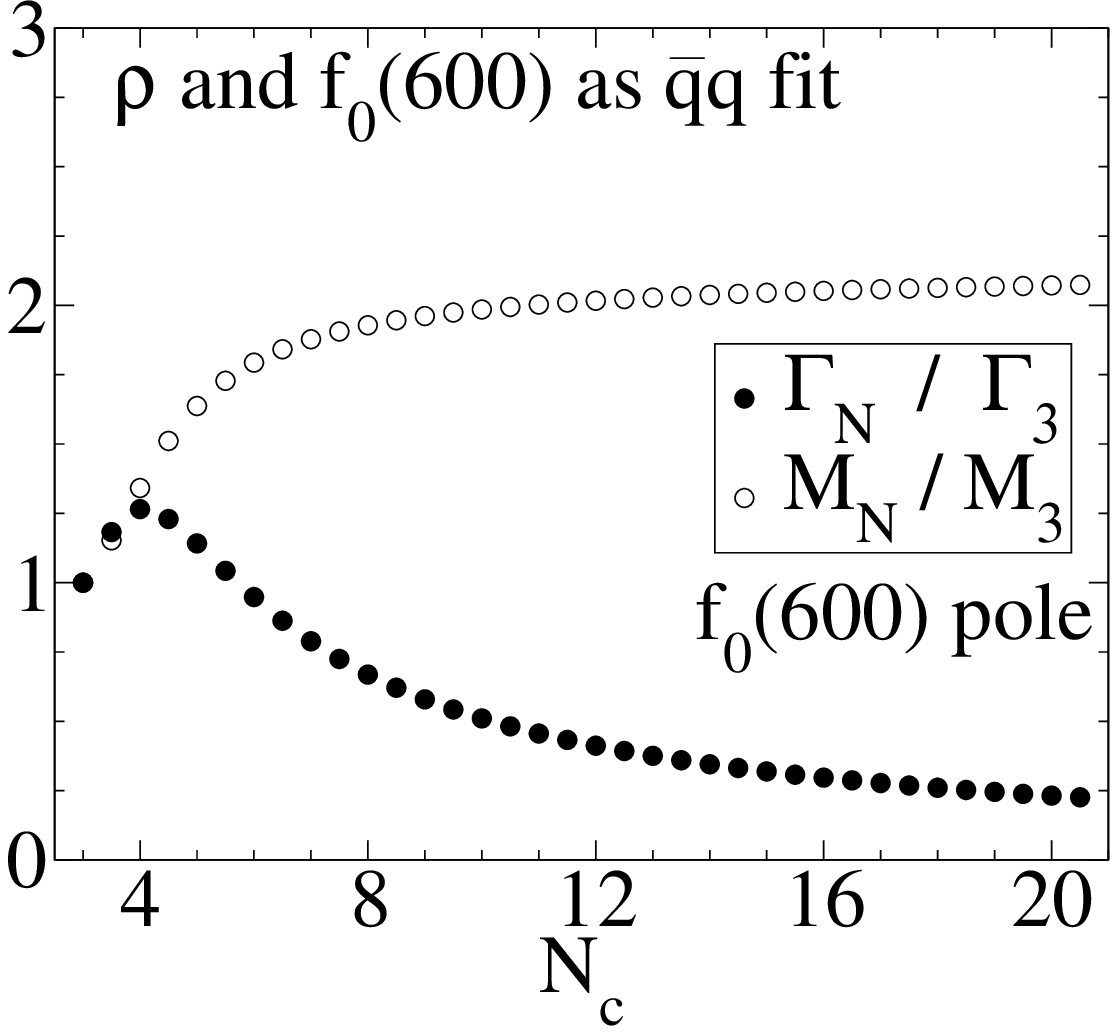}% Here is how to import EPS art
}
\caption{\rm Mass and width  $N_c$ behavior of the
$\rho$ and $f_0(600)$ from an $O(p^6)$ IAM data fit minimizing
{\it also} the $\bar \chi_{\bar{q}q}^2$: of the $\rho$ (Top). 
of the $f_0(600)$ instead of the $\rho$ (Center)
of both the $f_0(600)$ and $\rho$ (Bottom).}
\end{figure}
In summary, we have presented a method to determine 
quantitatively how
close the $N_c$ dependence
of a resonance pole is to a $\bar{q}q$ behavior. 
We have applied this measure to the poles generated 
in $\pi\pi$ scattering by unitarized Chiral Perturbation Theory,
which is the effective low energy theory of QCD and
reproduces systematically its large $N_c$ expansion.
The method is able to confirm the $O(p^4)$ qualitative 
results \cite{Pelaez:2003dy,Pelaez:2004xp}, identifying the $\rho$ as a $\bar{q}q$ state
and showing that the $f_0(600)$ is at odds 
with a dominant $\bar{q}q$ component.
We have extended the method to $O(p^6)$ confirming the stability
of our $O(p^4)$ conclusions, but also showing that a possible 
subdominant $\bar{q}q$ may originate around 1 GeV. This provides 
further support, based on the QCD $N_c$ dependence,
to some models that generate the $f_0(600)$ from
final state meson interactions, and
 locate a ``preexisting'' $\bar{q}q$ scalar nonet 
\cite{VanBeveren:1986ea,Oller:1997ti}
around 1 GeV. The methods presented here should be easily generalized
to investigate the nature of other dynamically generated 
mesons \cite{Dobado:2001rv} and baryons \cite{GomezNicola:2000wk}.

\vspace*{-.7cm}
\bibliography{apssamp}% Produces the bibliography via BibTeX.

\begin{thebibliography}{99}
\footnotesize
\vspace*{-.7cm}
%\cite{Aitala:2000xu}
\bibitem{Aitala:2000xu}
E.~M.~Aitala {\it et al.}  [E791 Collaboration],
%``Experimental evidence for a light and broad scalar resonance in D+ $\to$  pi- pi+ pi+ decay,''
Phys.\ Rev.\ Lett.\  {\bf 86}, 770 (2001)
%[arXiv:hep-ex/0007028].
%%CITATION = HEP-EX 0007028;%%
%\cite{Aitala:2002kr}
%\bibitem{Aitala:2002kr}
%E.~M.~Aitala {\it et al.}  [E791 Collaboration],
%``Dalitz plot analysis of the decay D+ $\to$ K- pi+ pi+ and study of the K  pi scalar amplitudes,''
Phys.\ Rev.\ Lett.\  {\bf 89}, 121801 (2002)
%[arXiv:hep-ex/0204018].
%%CITATION = HEP-EX 0204018;%%%
%\cite{Bediaga:2004bc}
%\bibitem{Bediaga:2004bc}
  I.~Bediaga and J.~M.~de Miranda,
%   ``Phase motion in the scalar low-mass pi pi amplitude in D+ --> pi- pi+  pi+
  %decay,''
  Phys.\ Lett.\ B {\bf 633}, 167 (2006)
%  [arXiv:hep-ex/0405019].
  %%CITATION = HEP-EX 0405019;%%
%\bibitem{Ablikim:2004qn}
M.~Ablikim {\it et al.}  [BES Collaboration],
%``The sigma pole in J/psi $\to$ omega pi+ pi-,''
Phys.\ Lett.\ B {\bf 598}, 149 (2004)
%[arXiv:hep-ex/0406038].
%%CITATION = HEP-EX 0406038;%%
%\cite{Bugg:2003kj}
%\bibitem{Bugg:2003kj}
D.~V.~Bugg,
%``Comments on the sigma and kappa,''
Phys.\ Lett.\ B {\bf 572}, 1 (2003)
[Erratum-ibid.\ B {\bf 595}, 556 (2004)],
%%CITATION = PHLTA,B572,1;%%
%\cite{Bugg:2004xu}
%\bibitem{Bugg:2004xu}
%D.~V.~Bugg,
%``Four sorts of meson,''
Phys.\ Rept.\  {\bf 397}, 257 (2004).
%%CITATION = PRPLC,397,257;%%

\bibitem{newsigma}
R.L. Jaffe, \PR{D15} 267 (1977); \PR{D15}, 281 (1977).
%\cite{Kaminski:1993zb}
%\bibitem{Kaminski:1993zb}
R.~Kaminski, L.~Lesniak and J.~P.~Maillet,
%``Relativistic effects in the scalar meson dynamics,''
Phys.\ Rev.\ D {\bf 50} (1994) 3145.
%[arXiv:hep-ph/9403264].
%%CITATION = HEP-PH 9403264;%%
%%CITATION = HEP-PH 9610325;%%
%\cite{Harada:1995dc}
%\bibitem{Harada:1995dc}
M.~Harada, F.~Sannino and J.~Schechter,
%``Simple Description of Pion-Pion Scattering to 1 GeV,''
Phys.\ Rev.\ D {\bf 54} (1996) 1991
%[arXiv:hep-ph/9511335].
%%CITATION = HEP-PH 9511335;%%
R.~Delbourgo and M.~D.~Scadron,
%``Dynamical Generation Of The SU(2) Linear Sigma Model,''
Mod.\ Phys.\ Lett.\ A {\bf 10} (1995) 251.
%[arXiv:hep-ph/9910242].
%%CITATION = HEP-PH 9910242;%%
%\cite{Ishida:1995xx}
%\bibitem{Ishida:1995xx}
S.~Ishida {\it et al.}, 
%M.~Ishida, H.~Takahashi, T.~Ishida, K.~Takamatsu and T.~Tsuru,
%``An Analysis of pi pi scattering phase shift and existence of sigma (555) particle,''
Prog.\ Theor.\ Phys.\  {\bf 95} (1996) 745;
Prog. Theor. Phys. 98,621 (1997).
%[arXiv:hep-ph/9610325].
%\cite{Tornqvist:1995ay}
%\bibitem{Tornqvist:1995ay}
N.~A.~Tornqvist and M.~Roos,
%``Resurrection of the Sigma Meson,''
Phys.\ Rev.\ Lett.\  {\bf 76} (1996) 1575.
%[arXiv:hep-ph/9511210].
%%CITATION = HEP-PH 9511210;%%
%\bibitem{kappa} 
%S. Ishida {\it et al}, 
D. Black {\it et al.},. 
\PR{D58}:054012,1998. 
%[arXiv:hep-ex/0106077].
%\cite{Colangelo:2001df}
%\bibitem{Colangelo:2001df}
  G.~Colangelo, J.~Gasser and H.~Leutwyler,
  %``pi pi scattering,''
  Nucl.\ Phys.\ B {\bf 603}, 125 (2001)
%  [arXiv:hep-ph/0103088].
  %%CITATION = HEP-PH 0103088;%%

%\cite{VanBeveren:1986ea}
\bibitem{VanBeveren:1986ea}
  E.~Van Beveren, {\it et al.} %T.~A.~Rijken, K.~Metzger, C.~Dullemond, G.~Rupp and J.~E.~Ribeiro,
  %``A Low Lying Scalar Meson Nonet In A Unitarized Meson Model,''
  Z.\ Phys.\ C {\bf 30}, 615 (1986)
  %%CITATION = ZEPYA,C30,615;%%
and 
%\cite{vanBeveren:2006ua}
%\bibitem{vanBeveren:2006ua}
%  E.~van Beveren, D.~V.~Bugg, F.~Kleefeld and G.~Rupp,
  %``The nature of sigma, kappa, a0(980) and f0(980),''
%  arXiv:
hep-ph/0606022.
  %%CITATION = HEP-PH 0606022;%%
%\cite{vanBeveren:2001kf}
%\bibitem{vanBeveren:2001kf}
E.~van Beveren and G.~Rupp,
%``Modified Breit-Wigner formula for mesonic resonances describing OZI  decays of confined q anti-q states and the light scalar mesons,''
Eur.\ Phys.\ J.\ C {\bf 22} (2001) 493,
%\cite{vanBeveren:2002vw}
%\bibitem{vanBeveren:2002vw}
%E.~van Beveren and G.~Rupp,
%``Scalar mesons within a model for all non-exotic mesons,''
hep-ph/0201006.
%%CITATION = HEP-PH 0201006;%%

%\cite{Dobado:1996ps}
\bibitem{Dobado:1996ps}
A.~Dobado and J.~R.~Pelaez,
%``A Global fit of pi pi and pi K elastic scattering in 
%ChPT with dispersion relations,''
Phys.\ Rev.\ D {\bf 47} (1993) 4883.
%[arXiv:hep-ph/9301276].
%%CITATION = HEP-PH 9301276;%%
%``The inverse amplitude method in Chiral Perturbation Theory,''
Phys.\ Rev.\ D {\bf 56} (1997) 3057.
%[arXiv:hep-ph/9604416].
%%CITATION = HEP-PH 9604416;%%
%\cite{Dobado:1992ha}


%\cite{Oller:1997ti}
\bibitem{Oller:1997ti}
J.~A.~Oller and E.~Oset,
%``Chiral symmetry amplitudes in the S-wave isoscalar and isovector  channels and the sigma, f0(980), a0(980) scalar mesons,''
Nucl.\ Phys.\ A {\bf 620} (1997) 438; 
[Erratum-ibid.\ A {\bf 652} (1999) 407]
%[arXiv:hep-ph/9702314].
%%CITATION = HEP-PH 9702314;%%
%\bibitem{Oller:1998zr}%\cite{Oller:1998zr}
%\bibitem{Oller:1998zr}
%J.~A.~Oller and E.~Oset,
%``N/D description of two meson amplitudes and chiral symmetry,''
Phys.\ Rev.\ D {\bf 60} (1999) 074023.
%[arXiv:hep-ph/9809337].
%%CITATION = HEP-PH 9809337;%%



%\cite{Oller:1997ng}
\bibitem{Oller:1997ng}
J.~A.~Oller, E.~Oset and J.~R.~Pelaez,
%``Non-perturbative approach to effective chiral Lagrangians and meson  interactions,''
Phys.\ Rev.\ Lett.\  {\bf 80} (1998) 3452;
%[arXiv:hep-ph/9803242].
%%CITATION = HEP-PH 9803242;%%
%\cite{Oller:1998hw}
%\bibitem{Oller:1998hw}
%J.~A.~Oller, E.~Oset and J.~R.~Pelaez,
%``Meson meson and meson baryon interactions in a chiral non-perturbative  approach,''
Phys.\ Rev.\ D {\bf 59} (1999) 074001
[Erratum-ibid.\ D {\bf 60} (1999) 09990],
%[arXiv:hep-ph/9804209].
%%CITATION = HEP-PH 9804209;%%
%\cite{Oller:1999ag}
%\bibitem{Oller:1999ag}
%J.~A.~Oller, E.~Oset and J.~R.~Pel\'aez,
%``The Phi $\to$ pi+ pi- decay within a chiral unitary approach,''
 and Phys.\ Rev.\ D {\bf 62} (2000) 114017.
%[arXiv:hep-ph/9911297].
%%CITATION = HEP-PH 9911297;%%
%\cite{Uehara:2002nv}
%\bibitem{Uehara:2002nv}
M.~Uehara,
%``Revisit to low mass scalar mesons via unitarized chiral perturbation  theory,''
%arXiv:
hep-ph/0204020.
%%CITATION = HEP-PH 0204020;%%



%\cite{Pelaez:2003dy}
\bibitem{Pelaez:2003dy}
  J.~R.~Pelaez,
  %``On the nature of light scalar mesons from their large N(c) behavior,''
  Phys.\ Rev.\ Lett.\  {\bf 92}, 102001 (2004)
  %[arXiv:hep-ph/0309292].
  %%CITATION = HEP-PH 0309292;%%

%\cite{Pelaez:2004xp
\bibitem{Pelaez:2004xp}
  J.~R.~Pelaez,
%   ``Light scalars as tetraquarks or two-meson states from large N(c) and
  %unitarized chiral perturbation theory,''
  Mod.\ Phys.\ Lett.\ A {\bf 19}, 2879 (2004)
  %[arXiv:hep-ph/0411107].
  %%CITATION = HEP-PH 0411107;%%


%\cite{'tHooft:1973jz}
\bibitem{'tHooft:1973jz}
G.~'t Hooft,
%``A Planar Diagram Theory For Strong Interactions,''
Nucl.\ Phys.\ B {\bf 72} (1974) 461.
%%CITATION = NUPHA,B72,461;%%
%\cite{Witten:1980sp}
%\bibitem{Witten:1980sp}
E.~Witten,
%``Large N Chiral Dynamics,''
Annals Phys.\  {\bf 128} (1980) 363.
%%CITATION = APNYA,128,363;%%

\bibitem{Jaffe} R. L. Jaffe, Proceedings of the Intl. Symposium
on Lepton and Photon Interactions at High Energies. Physikalisches Institut, University of Bonn (1981) . ISBN: 3-9800625-0-3 


\bibitem{chpt1}
S. Weinberg, Physica {\bf A96} (1979) 327.
%\cite{Gasser:1983yg}
%\bibitem{Gasser:1983yg}
J.~Gasser and H.~Leutwyler,
%``Chiral Perturbation Theory To One Loop,''
Annals Phys.\  {\bf 158} (1984) 142;
%%CITATION = APNYA,158,142;%%
%\cite{Gasser:1984gg}
%\bibitem{Gasser:1984gg}
%J.~Gasser and H.~Leutwyler,
%``Chiral Perturbation Theory: Expansions In The Mass Of The Strange Quark,''
Nucl.\ Phys.\ B {\bf 250} (1985) 465.
%%CITATION = NUPHA,B250,465;%%

%\cite{Truong:1988zp}
\bibitem{Truong:1988zp}
T.~N.~Truong,
%``Chiral Perturbation Theory And Final State Theorem,''
Phys.\ Rev.\ Lett.\  {\bf 61} (1988) 2526.
%%CITATION = PRLTA,61,2526;%%
\PRL{67}, (1991) 2260;
A. Dobado, M.J.Herrero and T.N. Truong, \PL{B235} (1990) 134.
%\bibitem{Dobado:1992ha}


%\cite{Guerrero:1998ei}
\bibitem{Guerrero:1998ei}
F.~Guerrero and J.~A.~Oller,
%``K anti-K scattering amplitude to one loop in chiral perturbation  
%theory, its unitarization and pion form factors,''
Nucl.\ Phys.\ B {\bf 537} (1999) 459
[Erratum-ibid.\ B {\bf 602} (2001) 641].
%[arXiv:hep-ph/9805334].
%%CITATION = HEP-PH 9805334;%%


%\cite{GomezNicola:2001as}
\bibitem{GomezNicola:2001as}
A.~G\'omez Nicola and J.~R.~Pel\'aez,
%``Meson meson scattering within one loop chiral perturbation theory 
%and  its unitarization,''.
Phys.\ Rev.\ D {\bf 65} (2002) 054009 and
%[arXiv:hep-ph/0109056].
%%CITATION = HEP-PH 0109056;%%
%\cite{Pelaez:2003xd}
%\bibitem{Pelaez:2003xd}
%J.~R.~Pelaez and A.~Gomez Nicola,
%``Light meson resonances from unitarized chiral perturbation theory,''
AIP Conf.\ Proc.\  {\bf 660} (2003) 102
[hep-ph/0301049].
%%CITATION = HEP-PH 0301049;%%




%\cite{Uehara:2003ax}
\bibitem{Uehara:2003ax}
M.~Uehara,
%``Can low mass scalar meson nonet survive in large N(c) limit?,''
%arXiv:
hep-ph/0308241,
%%CITATION = HEP-PH 0308241;%%
%\cite{Uehara:2004es}
%\bibitem{Uehara:2004es}
%M.~Uehara,
%``Large N(c) behavior of light scalar meson nonet revisited,''
hep-ph/0401037,
%%CITATION = HEP-PH 0401037;%%
%\cite{Uehara:2004ij}
%\bibitem{Uehara:2004ij}
%M.~Uehara,
%``A unitarized chiral approach to f0(980) and a0(980) states and nature of
%light scalar resonances,''
hep-ph/0404221.
%%CITATION = HEP-PH 0404221;%%


%\cite{Bijnens:1997vq}
\bibitem{Bijnens:1997vq}
  J.~Bijnens {\it et al.},%, G.~Colangelo, G.~Ecker, J.~Gasser and M.~E.~Sainio,
  %``Pion pion scattering at low energy,''
  Nucl.\ Phys.\ B {\bf 508}, 263 (1997)
%[Erratum-ibid.\ B {\bf 517}, 639 (1998)]
  %[arXiv:hep-ph/9707291].
  %%CITATION = HEP-PH 9707291;%%

%\cite{Amoros:1999qq}
\bibitem{Amoros:1999qq}
  G.~Amoros, J.~Bijnens and P.~Talavera,
  %``Low energy constants from K(l4) form-factors,''
  Phys.\ Lett.\ B {\bf 480}, 71 (2000)
  %[arXiv:hep-ph/9912398].
  %%CITATION = HEP-PH 9912398;%%


\bibitem{BijnensGasser} J. Bijnens, G. Colangelo and J. Gasser,
\NP{B427} (1994) 427.

%\cite{Girlanda:1997ed}
\bibitem{Girlanda:1997ed}
  L.~Girlanda, M.~Knecht, B.~Moussallam and J.~Stern,
%   ``Comment on the prediction of two-loop standard chiral perturbation  theory
  %for low-energy pi pi scattering,''
  Phys.\ Lett.\ B {\bf 409}, 461 (1997)
%  [arXiv:hep-ph/9703448].
  %%CITATION = HEP-PH 9703448;%%
%\cite{Nieves:1999zb}
%\bibitem{Nieves:1999zb}
  J.~Nieves and E.~Ruiz Arriola,
%   ``Error estimates for pi pi scattering threshold parameters in chiral
  %perturbation theory to two loops,''
  Eur.\ Phys.\ J.\ A {\bf 8}, 377 (2000)
%  [arXiv:hep-ph/9906437].
  %%CITATION = HEP-PH 9906437;%%

%\cite{Ecker:1988te}
\bibitem{Ecker:1988te}
  G.~Ecker, J.~Gasser, A.~Pich and E.~de Rafael,
  %``The Role Of Resonances In Chiral Perturbation Theory,''
  Nucl.\ Phys.\ B {\bf 321}, 311 (1989).
  %%CITATION = NUPHA,B321,311;%%
%\cite{Donoghue:1988ed}
%\bibitem{Donoghue:1988ed}
  J.~F.~Donoghue, C.~Ramirez and G.~Valencia,
%   ``THE SPECTRUM OF QCD AND CHIRAL LAGRANGIANS OF THE STRONG AND WEAK
  %INTERACTIONS,''
  Phys.\ Rev.\ D {\bf 39}, 1947 (1989).
  %%CITATION = PHRVA,D39,1947;%%



%\cite{Nieves:2001de}
\bibitem{Nieves:2001de}
J.~Nieves, M.~Pavon Valderrama and E.~Ruiz Arriola,
%``The inverse amplitude method in pi pi scattering in chiral perturbation  theory to two loops,''
Phys.\ Rev.\ D {\bf 65}, 036002 (2002)
%[arXiv:hep-ph/0109077].
%%CITATION = HEP-PH 0109077;%%



%\cite{Sun:2005uk}
\bibitem{Sun:2005uk}
  Z.~X.~Sun, {\it et al.} % L.~Y.~Xiao, Z.~G.~Xiao and H.~Q.~Zheng,
%   ``Model dependent analyses on the N(c) dependence of the sigma pole
  %trajectory,''
%  arXiv:
hep-ph/0503195.
  %%CITATION = HEP-PH 0503195;%%
%\cite{Pelaez:2005fd}
%\bibitem{Pelaez:2005fd}
  J.~R.~Pelaez,
  %``Comments on the large N(c) behavior of light scalars,''
%  arXiv:
hep-ph/0509284.
  %%CITATION = HEP-PH 0509284;%%





%\cite{Dobado:2001rv}
\bibitem{Dobado:2001rv}
  A.~Dobado and J.~R.~Pelaez,
  %``Chiral perturbation theory and the f2(1270) resonance,''
  Phys.\ Rev.\ D {\bf 65}, 077502 (2002)
%  [arXiv:hep-ph/0111140].
  %%CITATION = HEP-PH 0111140;%%
%\cite{Roca:2005nm}
%\bibitem{Roca:2005nm}
  L.~Roca, E.~Oset and J.~Singh,
  %``Low lying axial-vector mesons as dynamically generated resonances,''
  Phys.\ Rev.\ D {\bf 72}, 014002 (2005)
%  [arXiv:hep-ph/0503273].
  %%CITATION = HEP-PH 0503273;%%

%\cite{GomezNicola:2000wk}
\bibitem{GomezNicola:2000wk}
  A.~Gomez Nicola, J.~Nieves, J.~R.~Pelaez and E.~Ruiz Arriola,
%   ``Improved unitarized heavy baryon chiral perturbation theory for pi N
  %scattering,''
  Phys.\ Lett.\ B {\bf 486}, 77 (2000)
%  [arXiv:hep-ph/0006043].
  %%CITATION = HEP-PH 0006043;%%

\end{thebibliography}

\end{document}